\documentclass[prl,twocolumn,superscriptaddress]{revtex4}
\usepackage{psfig}

\begin{document}

\title{Comment on ``Nonmonotonic $d_{x^{2}-y^{2}}$ Superconducting
Order Parameter in Nd$_{2-x}$Ce$_x$CuO$_4$''}


\maketitle


In a recent Letter \cite{blumberg02} Blumberg {\it et al.} address
the symmetry of the superconducting gap in cuprate
superconductors. In particular in the electron-doped systems the
issue  is not yet settled, and not even the phase sensitive
experiments \cite{phase} arrive at a consistent conclusion.
Therefore more spectroscopic information is useful: in addition to
the B$_{1g}$ and B$_{2g}$ Raman spectra \cite{stadlober95}
Blumberg and coauthors measured the A$_{1g}$ component by using an
excitation energy of 1.9 eV. They obtain positions of the
A$_{1g}$, B$_{1g}$ and B$_{2g}$ pair breaking peaks at 45, 50 and
67~cm$^{-1}$, respectively, and conclude, from these positions
alone, that the superconducting order parameter has $d_{x^2-y^2}$
symmetry with a non-monotonic dependence on the azimuthal angle
$\phi$ (see Fig. 1b in Ref. \cite{blumberg02}).

\begin{figure}[t]
\centerline{\psfig{file=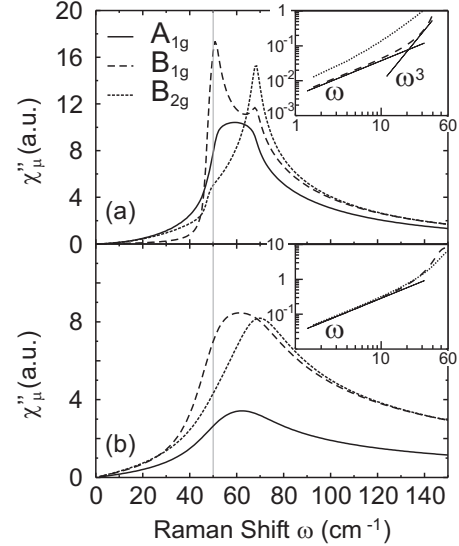,width=6cm}} \caption{Raman
response calculated with the gap function proposed in Ref.
\cite{blumberg02}. The best analytical approximation is given by
$\Delta(\phi) = \Delta_0  [\sin(2\phi)+a_1\sin(6\phi) +a_2
\sin(10\phi)]$ ($2\Delta_0 =67~{\rm cm^{-1}}$, $a_1=0.42$,
$a_2=0.17$) using the same definition of $\phi$ as in Ref.
\cite{blumberg02}. In this nomenclature the scattering amplitudes
read: $\gamma_{B_{1g}} \propto \sin(2\phi), \gamma_{B_{2g}}
\propto \cos(2\phi)$, and $\gamma_{A_{1g}} \propto -\cos(4\phi)$.
Neither for small (a) nor for large (b) damping the power laws, in
particular the $\omega^3$ dependence in $B_{1g}$ symmetry can be
observed any more for $\omega<20~{\rm cm^{-1}}$ (insets).}
\label{Fig}
\end{figure}

In this Comment we show that the basis for this conclusion is
insufficient. This becomes already clear by just following the
qualitative arguments of the authors: since the Raman scattering
amplitudes $\gamma_{\mu}(\phi)$ of all symmetries $\mu$ are finite
at the maximum $\Delta_0$ of the proposed gap function, the
spectra in {\em all} symmetries will exhibit structures at the
same energy $2\Delta_0$ as opposed to what is observed
\cite{blumberg02,stadlober95}. In addition, if $\Delta(\phi)$ has
components up to $\sin(10\phi)$ (see caption of Fig.~1) as
proposed in Ref.~\cite{blumberg02} it is hard to probe them with
$\gamma_{\mu} \propto \sin(2\phi)$ without applying a model.
Therefore, we calculated  the Raman response explicitly.
\cite{theory} For a small phenomenological damping
$\Gamma/\Delta_0=0.04$ (Fig. \ref{Fig}~a) both the $B_{1g}$ and
the $B_{2g}$ spectra exhibit two structures at around 50 and
67~${\rm cm^{-1}}$, and the $A_{1g}$ peak is at approximately 60
and not at 45~${\rm cm^{-1}}$ as in the experiment. Even for this
small damping all spectra are linear (see inset of
Fig.~\ref{Fig}~a) up to approximately 20~${\rm cm^{-1}}$
\cite{theory}. A larger damping $\Gamma/\Delta_0=0.30$ (Fig.
\ref{Fig}~b) not only smears out the double peak structure but
also completely kills the power laws characteristic for $d$-wave
pairing for this type of gap.
In general, neither the {\bf k}~dependence nor the magnitude of
the gap can be derived from the peak positions alone. Rather an
appropriate model and the complete study of the low-temperature
and low-frequency power laws are required. Then, constraints for
the gap similar to those from the specific heat or the magnetic
penetration depth could be obtained.

Here, it is indeed the magnetic penetration depth \cite{lambda}
$\lambda(T)$ which is in direct conflict with the proposed form of
the gap. The temperature dependence of $\lambda(T)$ can be readily
calculated from the functional dependence $\Delta(\phi)$. No
agreement with the data \cite{lambda} at any doping can be
achieved. Supposed the nonmonotonic gap would be realized an
analysis with a monotonic one \cite{lambda} would lead to
$\Delta_0 \simeq 9k_BT_c$ in spite of the restricted phase space
around the node.\\

In conclusion, the functional dependence of the gap proposed in
Ref. \cite{blumberg02} is neither sufficiently supported by the
Raman results nor compatible with the magnetic pentration depth.
In spite of that, an anisotropic $s$-wave as suggested earlier
\cite{stadlober95} is probably not the full story either at least
not in the entire doping range. Therefore the issue of the
superconducting gap in the electron-doped systems cannot at all be
considered solved by now.\\

F. Venturini and R. Hackl, Walther Meissner Institute, Bavarian
Academy of Sciences, 85748 Garching, Germany

U. Michelucci, EKM, Universit\"at Augsburg, 86135 Augsburg,
Germany



\end{document}